\documentclass[conference,compsoc]{IEEEtran}

\ifCLASSOPTIONcompsoc
  \usepackage[nocompress]{cite}
\else
  \usepackage{cite}
\fi

\input{meta/packages}

\newcommand{\footurl}[1]{{\footnote{\url{#1}}}}

\newcommand{\new}[1]{#1}

\newcommand{\checkNum}[1]{#1}

\newcommand{\toolUsage}{\textit{go-geiger}}
\newcommand{\toolSA}{\textit{go-safer}}
\newcommand{\unsafe}{\textit{unsafe}}

\newcommand{\projsAnalyzed}{\checkNum{343}}
\newcommand{\initalProjs}{\checkNum{500}}
\newcommand{\withoutModules}{\checkNum{150}}
\newcommand{\notCompiled}{\checkNum{7}}
\newcommand{\packagesAnalyzed}{\checkNum{62,025}}
\newcommand{\packagesAnalyzedRounded}{\checkNum{62,000}}

\newcommand{\numberPRs}{\checkNum{14}}
\newcommand{\numberPRsMerged}{\checkNum{10}}
\newcommand{\numberBugsFixed}{\checkNum{60}}

\newcommand{\numberCodeSnippets}{\checkNum{1,400}}

\newcommand{\percentageProjectsWithUnsafe}{\checkNum{38\%}}

\newcommand{\percentageProjectsAndDependenciesUnsafe}{\checkNum{91\%}}

\newcommand{\averageUnsafeImportDepth}{\checkNum{3.08}}
\newcommand{\stdUnsafeImportDepth}{\checkNum{1.62}}
\newcommand{\averageGeneralImportDepth}{\checkNum{3.04}}

\newcommand{\levelOneImportedUnsafePackagesCount}{\checkNum{569}}
\newcommand{\levelOneImportedUnsafePackagesShare}{\checkNum{3.63\%}}
\newcommand{\levelOneImportedPackagesCount}{\checkNum{8,952}}

\begin{document}
\title{Uncovering the Hidden Dangers: \\Finding Unsafe Go Code in the Wild}

\author{
\IEEEauthorblockN{
Johannes Lauinger\IEEEauthorrefmark{2},
Lars Baumgärtner\IEEEauthorrefmark{1}, 
Anna-Katharina Wickert\IEEEauthorrefmark{1},
Mira Mezini\IEEEauthorrefmark{1}
}
\IEEEauthorblockA{
Technische Universität Darmstadt, D-64289 Darmstadt, Germany\\
}
\IEEEauthorblockA{\IEEEauthorrefmark{1}
E-mail: \{baumgaertner, wickert, mezini\}@cs.tu-darmstadt.de \\
}
\IEEEauthorblockA{\IEEEauthorrefmark{2}
E-mail: jlauinger@seemoo.tu-darmstadt.de \\
}
}

\IEEEoverridecommandlockouts
\IEEEpubid{\makebox[\columnwidth]{\copyright2020 IEEE. \hfill} \hspace{\columnsep}\makebox[\columnwidth]{ }}

\maketitle

\IEEEpubidadjcol
\begin{abstract}
The Go programming language aims to provide memory and thread safety through measures such as automated memory management with garbage collection and a strict type system.
However, it also offers a way of circumventing this safety net through the use of the \unsafe{} package.
While there are legitimate use cases for \unsafe{}, developers must exercise caution to avoid introducing vulnerabilities like buffer overflows or memory corruption in general.
In this work, we present \toolUsage{}, a novel tool for Go developers to quantify \unsafe{} usages in a project's source code and all of its dependencies.
Using \toolUsage{}, we conducted a study on the usage of \unsafe{} in the top 500 most popular open-source Go projects on GitHub, including a manual analysis of \numberCodeSnippets{} code samples on how \unsafe{} is used.
From the projects using Go's module system, %
\percentageProjectsWithUnsafe{} %
directly contain at least one \unsafe{} usage,
and \percentageProjectsAndDependenciesUnsafe{} contain at least one \unsafe{} usage in the project itself or one of its transitive dependencies.
Based on the usage patterns found, we present possible exploit vectors in different scenarios. 
Finally, we present \toolSA{}, a novel static analysis tool to identify dangerous and common usage patterns that were previously undetected with existing tools.

\end{abstract}


\IEEEpeerreviewmaketitle

\section{Introduction}
\label{sec:intro}

Programming languages with direct memory access through pointers, such as C/C++, suffer from the dangers of memory corruption, including buffer overflows \cite{alnaeli2017, larochelle2001} or \textit{use-after-free} of pointers.
Microsoft, e.g., reports that memory safety accounts for around 70\% of all their bugs\footnote{\url{https://msrc-blog.microsoft.com/2019/07/16/a-proactive-approach-to-more-secure-code/}}. 
To avoid these dangers, many programming languages, such as Java, Rust, Nim, or Google's Go, use automatic memory management and prevent using low-level memory details like pointers in favor of managed object references.
Thus, these languages are memory safe, eliminating most memory corruption bugs. 
However, there are valid use cases for such low-level features.
Safe languages therefore provide, to varying degrees, escape hatches to perform potentially unsafe operations.
Escape hatches may be used for optimization purposes, to directly access hardware, to use the foreign function interface (FFI), to access external libraries, or to circumvent limitations of the programming language. 

However, escape hatches may have severe consequences, e.g., they may introduce vulnerabilities.
This is especially problematic when \unsafe{} code blocks are introduced through third-party libraries, and thus \new{are} not directly obvious to the application developer. 
Indeed, a recent study shows that unsafe code blocks in Rust are often introduced through third-party libraries~\cite{evans2020}. 
\new{Therefore}, security analysts, developers, and administrators need efficient tools to quickly evaluate potential risks in their code base but also the risks introduced by code from others.

In this paper, we investigate Go and the usage of \unsafe{} code blocks within its most popular software projects. 
We developed two specific tools for developers and security analysts.
The first one, called \toolUsage{} (Section~\ref{sec:appr:toolUsage}) analyzes a project including its dependencies for locating usages of the \unsafe{} API and scoring \unsafe{} usages in Go projects and their dependencies. 
It is intended to give a general overview of \unsafe{} usages in a project. %

As \unsafe{} usages are benign when used correctly, safe usages of \unsafe{} exist.
\new{However, we identified several commonly used \unsafe{} patterns, e.g., to cast slices and structs, which can break memory safety mechanisms.
They introduce potential vulnerabilities, e.g., by allowing access to additional memory regions. 
We provide insights into the dangers and possible exploit vectors to these patterns, indicating the severe nature of these bugs leading to information leaks or code execution (Section~\ref{sec:appr:vulnerabilites}).

While the Go tool chain provides a linter, called \textit{go vet}, covering invalid \unsafe{} pointer conversions, 
the linter fails to flag the potentially insecure usages. 
Thus, to support developers we implemented a second tool \toolSA{} (Section~\ref{sec:appr:toolSA}) covering two types of those.}

With the help of \toolUsage{}, we performed a quantitative evaluation of the top \initalProjs{} most-starred Go projects on GitHub to see how often \unsafe{} is used in the wild (Section~\ref{sec:eval:unsafewild}). 
Including their dependencies, we analyzed more than \packagesAnalyzedRounded{} individual packages. %
We found that \percentageProjectsWithUnsafe{} of projects contain \unsafe{} usages in their direct application code, and \percentageProjectsAndDependenciesUnsafe{} of
projects contain \unsafe{} usages either in first-party or imported third-party libraries.

We also created a novel data set with \checkNum{1,400} labeled occurrences of \unsafe{}, providing insights into the motivation for introducing \unsafe{} in the source code in the first place (Section~\ref{sec:eval:labeledData}). 
\new{Finally, we used \toolSA{} to find instances of our identified dangerous usage patterns within the data set.}
So far, in the course of this work we submitted \numberPRs{} pull requests to analyzed projects and libraries, fixing over \numberBugsFixed{} individual potentially dangerous \unsafe{} usages \new{(Section~\ref{sec:discussion})}. %

In this paper, we make the following contributions:
\begin{itemize}
\item \toolUsage{}, a first-of-its-kind tool for detecting and scoring \unsafe{} usages in Go projects and their dependencies,
\item a novel static code analysis tool, \toolSA{}, to aid in identifying potentially problematic \unsafe{} usage patterns that were previously uncaught with existing tools,
\item a quantitative evaluation on the usage of \unsafe{} in \projsAnalyzed{} top-starred Go projects on GitHub,
\item a novel data set with \checkNum{1,400} labeled occurrences of \unsafe{}, providing insights into what is being used in real-world Go projects and for what purpose, and
\item evidence on how to exploit \unsafe{} usages in the wild.
\end{itemize}

\section{\new{Scanning for Usages of Go's \unsafe{} Package}}

In this section, we give a brief introduction to \unsafe{} in Go and then present %
\new{our novel standalone tool \toolUsage{} to identify \unsafe{} usages in a project and its dependencies. 
Thus, it supports auditing a project and perhaps selecting dependencies more carefully.}

\subsection{\new{Go's \unsafe{} Package}}

The Go programming language, like other memory-safe languages, provides an \unsafe{}~package\footnote{\url{https://golang.org/pkg/unsafe}}, which offers 
(a) the functions \textit{Sizeof}, \textit{Alignof}, and \textit{Offsetof} that are evaluated at compile time and provide access into memory alignment details of Go data types that are otherwise inaccessible, %
and (b) a pointer type, \textit{unsafe.Pointer}, that allows developers to circumvent restrictions of regular pointer types.

One can cast any pointer to/from \textit{unsafe.Pointer}, thus enabling casts between completely arbitrary types, as illustrated in  
Listing~\ref{lst:unsafe-ex-in-place-cast}.
In this example, \textit{in.Items} is assigned to a new type (\textit{out.Items}) in Line~3 without reallocation for efficiency reasons. %
Furthermore, casts between \textit{unsafe.Pointer} and \textit{uintptr} are also enabled, mainly for pointer arithmetic.
A \textit{uintptr} is an integer type with a length sufficient to store memory addresses. 
However, it is not a pointer type, hence, not treated as a reference.
Listing~\ref{lst:unsafe-ex-escape-analysis} presents an example of casts involving \textit{uintptr}. 
In Line~2, the \textit{unsafe.Pointer} is converted to \textit{uintptr}.
Thus, the memory address is stored within a non-reference type.
Hence, the back-conversion in Line~3 causes the \textit{unsafe.Pointer} to be hidden from the \textit{escape analysis (EA)} which Go's garbage collector uses 
to determine whether a pointer is local to a function and can be stored in the corresponding stack frame, 
or whether it can \textit{escape} the function and needs to be stored on the heap \cite{wang2020}. 
Storing the address of a pointer in a variable of \textit{uintptr} type and then converting it back causes the \textit{EA} to miss the chain of references to the underlying value in memory. 
Therefore, Go will assume a value does not escape when it actually does, and may place it on the stack.
Correctly used it can improve efficiency because deallocation is faster on the stack than on the heap~\cite{wang2020}.
However, used incorrectly it can cause security problems as shown later in Section~\ref{sec:appr:vulnerabilites}.

\begin{lstlisting}[language=Golang, label=lst:unsafe-ex-in-place-cast, caption=In-place cast using the \unsafe{} package
from the Kubernetes \textit{k8s.io/apiserver} module with minor changes.
,float, belowskip=-1.5em, aboveskip=0em]
func autoConvert(in *PolicyList, out *audit.PolicyList) error {
	// [...]
	out.Items = *(*[]audit.Policy)(unsafe.Pointer(&in.Items))
	return nil
}
\end{lstlisting}

\begin{lstlisting}[language=Golang, label=lst:unsafe-ex-escape-analysis, caption=Hiding a value from escape analysis from the \textit{modern-go/reflect2} module.
, float, belowskip=-1.5em]
func NoEscape(p unsafe.Pointer) unsafe.Pointer {
	x := uintptr(p)
	return unsafe.Pointer(x ^ 0)
}
\end{lstlisting}

\subsection{\new{\toolUsage{}: Identification of Unsafe Usages}}
\label{sec:appr:toolUsage}

\new{To identify and quantify usages of \unsafe{} in a Go project and its dependencies, we developed \toolUsage{}\footnote{\url{https://github.com/jlauinger/go-geiger}}.}
Its development was inspired by \textit{cargo geiger}\footnote{\url{https://github.com/rust-secure-code/cargo-geiger}}, a similar tool for detecting unsafe code blocks in Rust programs.

\begin{figure*}[htp!]
\centering
    \includegraphics[width=0.8\textwidth]{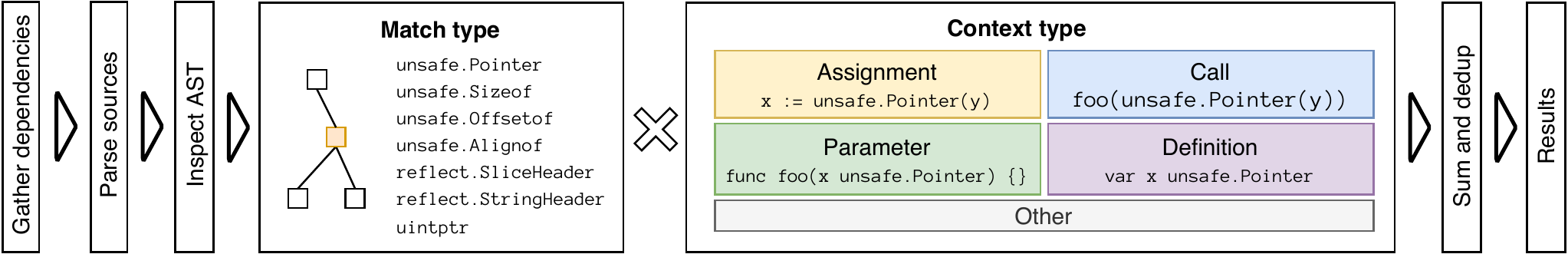}
    \caption{Architecture of \toolUsage{} tool to detect \unsafe{} usages}
    \label{fig:geiger-architecture}
    \vspace{-10pt}
\end{figure*}

Figure~\ref{fig:geiger-architecture} shows an overview of the architecture of \toolUsage{}.
We use the standard parsing infrastructure provided by Go to identify and parse packages including their dependencies based on user input.
Then, we analyze the AST, %
which enables us to identify different usages of \unsafe{} and their context as described in the next paragraph.
Finally, we arrange the packages requested for analysis and their dependencies in a tree structure, sum up \unsafe{} usages for each package individually, and calculate a cumulative score including dependencies.
We perform a deduplication if the same package is transitively imported more than once.
The \unsafe{} dependency tree, usage counts, as well as identified code snippets, are presented to the user.

We detect all usages of methods and fields from the \textit{unsafe} package, specifically: \textit{Pointer}, \textit{Sizeof}, \textit{Offsetof}, and \textit{Alignof}.
Furthermore, because they often are used in unsafe operations, we also count occurrences of \textit{SliceHeader} and \textit{StringHeader} from the \textit{reflect} package, and \textit{uintptr}.
All of these usages are referred to as \unsafe{} usages in this paper.
Additionally, we determine the context in which the \unsafe{} usage is found, i.e., 
the type of statement that includes the \unsafe{} usage.
In \toolUsage{} we distinguish between assignments (including definitions of composite literals and return statements), calls to functions, function parameter declarations, general variable definitions, or other not further specified usages.
We determine the context by looking up in the AST starting from the node representing the \unsafe{} usage, and identifying the type of the parent node.

\section{\new{Identifying Insecure Usages of \unsafe{}}}

\new{In this section, we present problematic code snippets including exploit information that we identified.
Further, we introduce our linter \toolSA{} to identify two known potentially dangerous \unsafe{} patterns for slice and struct casts.}

\subsection{\new{Potential Usage and Security Problems}}

\label{sec:appr:vulnerabilites}

In the following, we discuss potential threat models and exploit vectors against real-world \unsafe{} Go code.
We present a code pattern in Listing~\ref{lst:string-to-bytes} that is very common in popular open-source Go projects (cf. Section~\ref{sec:eval}).
It is used to convert a string to a byte slice without copying the data.
As in Go strings essentially are read-only byte slices, this is commonly done by projects to increase efficiency of serialization operations. %
Internally, each slice is represented by a data structure that contains its current length, allocated capacity, and memory address of the actual underlying data array.
The \textit{reflect} header structures provide access to this internal representation.
In Listing~\ref{lst:string-to-bytes} this conversion is done in Lines~2, 3, and~8 respectively.
First, an \textit{unsafe.Pointer} is used to convert a string to a \textit{reflect.StringHeader} type.
Then, a \textit{reflect.SliceHeader} instance is created and its fields are filled by copying the respective values from the string header.
Finally, the slice header object is converted into a slice of type \textit{[]byte}.

\begin{lstlisting}[language=Golang, label=lst:string-to-bytes, caption=Conversion from string to bytes using \unsafe{}, float, belowskip=-1.5em, aboveskip=0em]
func StringToBytes(s string) []byte {
	strHeader := (*reflect.StringHeader)(unsafe.Pointer(&s))
	bytesHeader := reflect.SliceHeader{
		Data: strHeader.Data,
		Cap:  strHeader.Len,
		Len:  strHeader.Len,
	}
	return *(*[]byte)(unsafe.Pointer(&bytesHeader))
}
\end{lstlisting}

\subsubsection*{Implicit Read-Only}

The conversion pattern shown in Listing~\ref{lst:string-to-bytes} is efficient as it directly casts between \textit{string} and \textit{[]byte} in-place. %
Using \textit{bytes := ([]byte)(s)} for the conversion would make the compiler allocate new memory for the slice header as well as the underlying data array.
However, the direct cast creates an implicitly read-only byte slice that can cause problems, as described in the following.
The Go compiler will place strings into a constant data section of the resulting binary file.
Therefore, when the binary is loaded into memory, the \textit{Data} field of the string header may contain an address that is located on a read-only memory page.
Hence, strings in Go are immutable by design and mutating a string causes a compiler error. %
However, when casting a string to a \textit{[]byte} slice in-place, the resulting slice loses the explicit read-only property, and thus, the compiler will not complain about mutating this slice although the program will crash if done so.

\subsubsection*{Garbage Collector Race}

Go uses a concurrent mark-and-sweep \textit{garbage collector (GC)} to free unused memory~\cite{sibiryov2017}.
It is triggered either by a certain increase of heap memory usage or after a fixed time. %
The GC treats pointer types, \textit{unsafe.Pointer} values, and slice/string headers as references and will mark them as still in use. %
Importantly, string/slice headers that are created manually as well as \textit{uintptr} values are not treated as references.
The last point, although documented briefly in the \textit{unsafe} package, is a major pitfall.
Casting a \textit{uintptr} variable back to a pointer type creates a potentially dangling pointer because the memory at that address might have already been freed if the GC was triggered right before the conversion.

\begin{figure}[!t]
    \centering
    \includegraphics[width=0.3\textwidth]{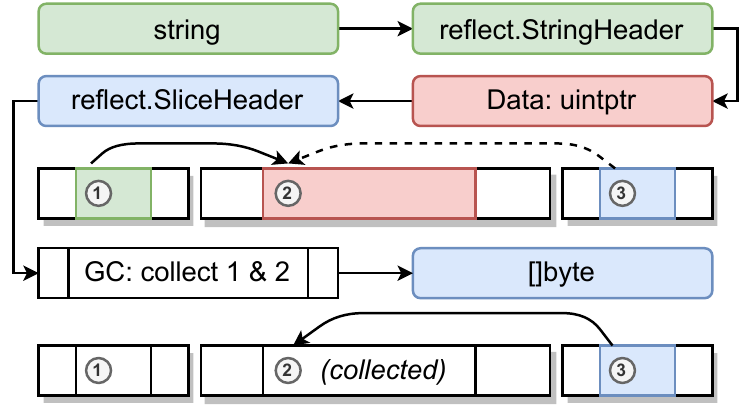}
    \caption{GC race and escape analysis flaw}
    \label{fig:gcrace-vuln}
    \vspace{-9pt}
\end{figure}

Although not directly obvious, Listing~\ref{lst:string-to-bytes} contains such a condition.
Because the \textit{reflect.SliceHeader} value is created as a composite literal instead of being derived from an actual slice value, its \textit{Data} field is not treated as a reference if the GC runs between Lines~3 and~8. 
Thus, the underlying data array of the \textit{[]byte} slice produced by the conversion might have already been collected.
This creates a potential \textit{use-after-free} or buffer reuse condition that, even worse, is triggered non-deterministically when the GC runs at just the "right" time.
Therefore, this race condition can crash the program, create an information leak, or even potentially lead to code execution.
Figure~\ref{fig:gcrace-vuln} shows a visualization of the casting process that leads to the problems described here.
The original slice is being cast to a string via some intermediate representations.
The slice header is shown in green (at memory position 1), while the underlying data array (memory position 2) is shown in red.
When the resulting string header (shown in blue at memory position 3) is created, it only has a weak reference to the data, and when the GC runs before converting it to the final string value, the data is already freed.

\subsubsection*{Escape Analysis Flaw}

A third problem found in Listing~\ref{lst:string-to-bytes} is that the \textit{escape analysis (EA)} algorithm can not infer a connection between the string parameter \textit{s} and the resulting byte slice.
Although they use the same underlying data array, the EA misses this due to the fact that the intermediate representation as a \textit{uintptr} is not treated as a reference type.
This can cause undefined behavior if the returned value from the casting function is used incorrectly.
Listing~\ref{lst:escape-analysis} shows a program that uses the conversion function presented earlier~(Listing~\ref{lst:string-to-bytes}).
In the \textit{main} function, \textit{GetBytes} is called~(Line~2), which creates a string and turns it into a byte slice using the conversion function.
Within the \textit{GetBytes} function, we create the string using a \textit{bufio} reader similarly to if it were user-provided input.
After the cast, \textit{GetBytes} prints the resulting bytes~(Line~10) and returns them to \textit{main}, which also prints the bytes~(Line~3).
Although one might assume that both print statements result in the same string to be displayed, the second one in \textit{main} fails and prints invalid data.

\begin{lstlisting}[language=Golang, label=lst:escape-analysis, caption=Escape analysis flaw example, float, belowskip=-1.5em, aboveskip=1em]
func main() {
	bytesResult := GetBytes()
	fmt.Printf("main: %s\n", bytesResult)
}

func GetBytes() []byte {
	reader := bufio.NewReader(strings.NewReader("abcdefgh"))
	s, _ := reader.ReadString('\n')
	out := StringToBytes(s)
	fmt.Printf("GetBytes: %s\n", out)
	return out
}
\end{lstlisting}

Because the string \textit{s} is allocated in \textit{GetBytes} the Go EA is triggered. %
It concludes that \textit{s} is passed to \textit{StringToBytes} and the EA transitively looks into that function.
Here, it fails to connect \textit{s} to the returned byte slice as described previously.
Therefore, the EA concludes that \textit{s} does not escape in \textit{StringToBytes}.
As it is not used after the call in \textit{GetBytes}, the EA algorithm incorrectly assumes that it does not escape at all and places \textit{s} on the stack.
When \textit{GetBytes} prints the resulting slice, the data is still valid and the correct data is printed, but once the function returns to \textit{main}, its stack is destroyed.
Thus, \textit{bytesResult} (Line~2) is now a dangling pointer into the former stack of \textit{GetBytes} and, therefore, printing data from an invalid memory region.

\subsubsection*{Code Execution}

To show the severity of the issues identified above and that they are not just of theoretical nature, e.g., resulting in simple program crashes, we created a proof of concept for a code execution exploit using \textit{Return Oriented Programming (ROP)} on a vulnerable \unsafe{} usage. %
The sample incorrectly casts an array on the stack into a slice without constricting it to the proper length.
This vulnerability causes a buffer overflow which we use to overwrite the stored return address on the stack, thus, changing the control flow of the program. 
Since Go programs are typically statically linked with a big runtime, there is a large number of ROP-gadgets available within the binary itself. 
We use gadgets to set register values and dispatch to system calls. 
Using the \textit{mprotect} syscall, we set both the writable and executable permission bits on a memory page that is mapped to the program, and store an exploit payload provided via standard input there using the \textit{read} syscall. 
Finally, we jump to this payload and execute it using a final ROP-gadget to open a shell.
An in-depth discussion of the exploit would go beyond the scope of this paper and exceed the space available to present our research.
Therefore, we made it available online\footnote{\url{https://github.com/jlauinger/go-unsafepointer-poc}\label{fn:poc}} \new{together with five other proof-of-concept exploits}. 
\new{Furthermore, we published an in-depth write-up about exploiting \unsafe{} issues in Go as a series of blog articles\footurl{https://dev.to/jlauinger/exploitation-exercise-with-unsafe-pointer-in-go-information-leak-part-1-1kga}.}

\subsection{\new{\toolSA{}: Finding Potentially Insecure Usages}}

\label{sec:appr:toolSA}

\new{We designed \toolSA{}\footnote{\url{https://github.com/jlauinger/go-safer}} to automatically give advice for some of the \unsafe{} usage patterns introduced in the previous section.
It is meant for assistance during manual audits and also for integration in build chains during development.
Avoiding the patterns that \toolSA{} detects prevents the garbage collector race and escape analysis flaw vulnerabilities that we discussed in Section~\ref{sec:appr:vulnerabilites}.
They are not covered by existing linters such as \textit{go vet}.
We found instances of these \unsafe{} code patterns through the usage of \toolSA{} in real-world code (cf. Section~\ref{sec:eval}).
}

\begin{figure}[t!] %
    \centering
    \includegraphics[width=0.35\textwidth]{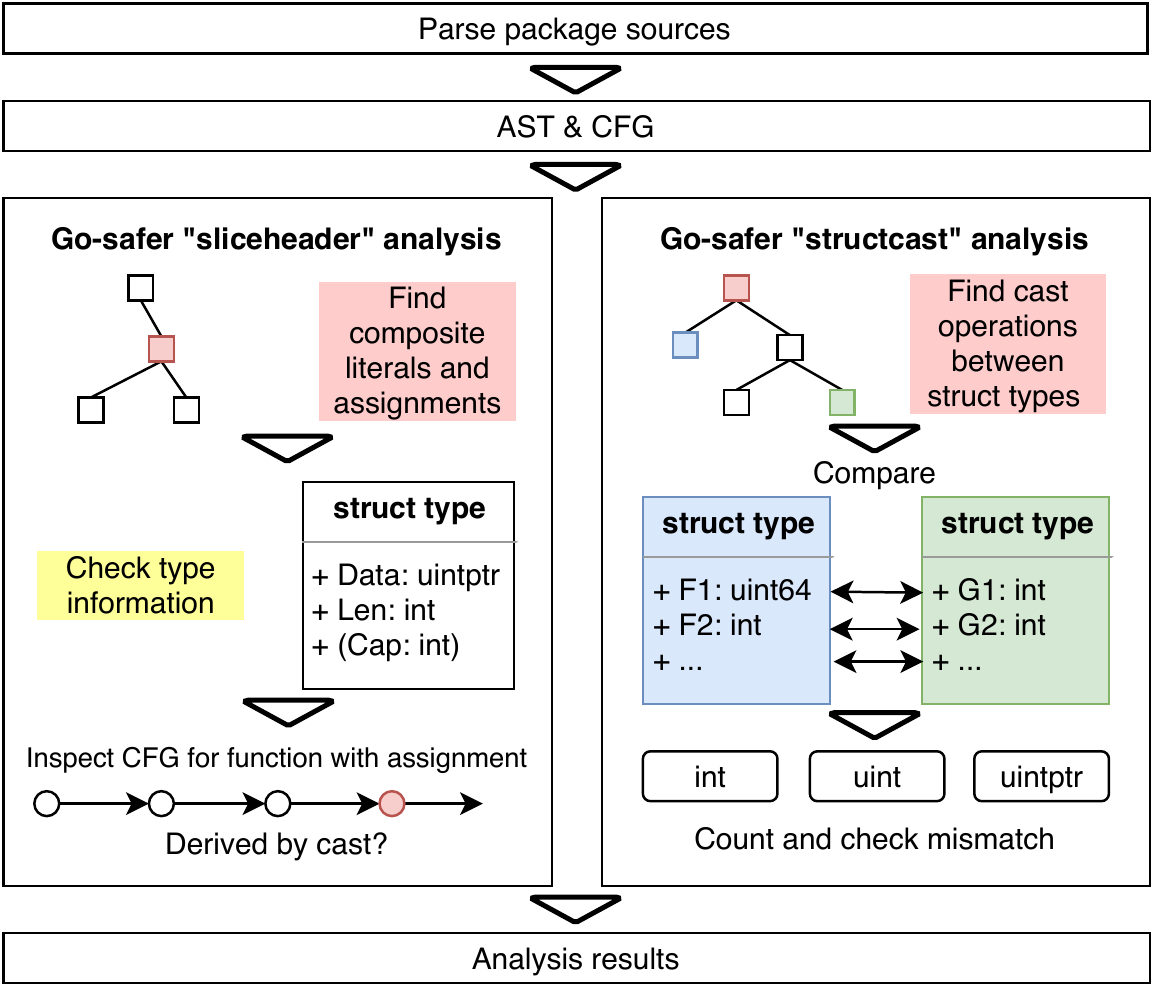}
    \caption{Architecture of \toolSA{} static code analysis tool}
    \label{fig:safer-architecture}
\end{figure}

Figure~\ref{fig:safer-architecture} shows an overview of the architecture of \toolSA{}.
First, it uses \textit{go vet} to build a list of packages to be analyzed and parses their sources.
Then, a number of static code analyzers, called \textit{passes}, run.
Our analyses depend on existing passes to acquire the abstract syntax tree (AST) and control flow graph (CFG).
Two separate analyses are run by \toolSA{}: the \textit{sliceheader} and the \textit{structcast} passes. %

The \textit{sliceheader} pass discovers incorrect string and slice casts as shown in Listing~\ref{lst:string-to-bytes}.
It finds composite literals and assignments in the AST.
Then, for each it checks whether the type of the %
receiver is \textit{reflect.StringHeader}, \textit{reflect.SliceHeader}, or some derived type with the same signature.
For assignments, the analysis pass then finds the last node in the CFG where the receiver object's value is defined, and checks if it is derived correctly by casting a string/slice.
If \toolSA{} can not infer with certainty that the assignment receiver object was created by a cast, a warning is issued.

The \textit{structcast} pass discovers instances of in-place casts between different struct types that include architecture-dependent field sizes. 
This can create a security risk when ported to other platforms because \unsafe{} casts can lead to misaligned fields, and thus, memory access outside a value's bounds on some platforms, allowing the same exploit vectors as a buffer overflow does.
The pass finds struct cast instances that involve \textit{unsafe.Pointer} in the AST.
Then, it compares the struct types and checks if they contain an unequal amount of fields with types \textit{int}, \textit{uint}, or \textit{uintptr}, which are the architecture-dependent types supported by Go.
If the numbers do not match, \toolSA{} issues a warning.

\section{A Study of Go's \unsafe{} Usages in the Wild}
\label{sec:eval}

We designed and performed a study of Go \unsafe{} usage to answer the following research questions:

\begin{enumerate}[leftmargin=*,label={RQ\arabic*}]
    \item How prevalent is \unsafe{} in Go projects? \label{rq:prevalApp}
    \item How deep are \unsafe{} code packages buried in the dependency tree? \label{rq:depsDepth}
    \item Which \unsafe{} keywords are used most? \label{rq:distTypes}
    \item Which \unsafe{} operations are used in practice, and for what purpose? \label{rq:purpose}
\end{enumerate}

In the following, we first describe our evaluation data set and then provide in-depth analyses of \unsafe{} usage in the wild using \toolUsage{}.
Our evaluation scripts as well as the results are available online\footnote{\url{https://github.com/stg-tud/unsafe_go_study_results}}.

\begin{figure*}[!t]
    \centering
    \includegraphics[width=.8\textwidth]{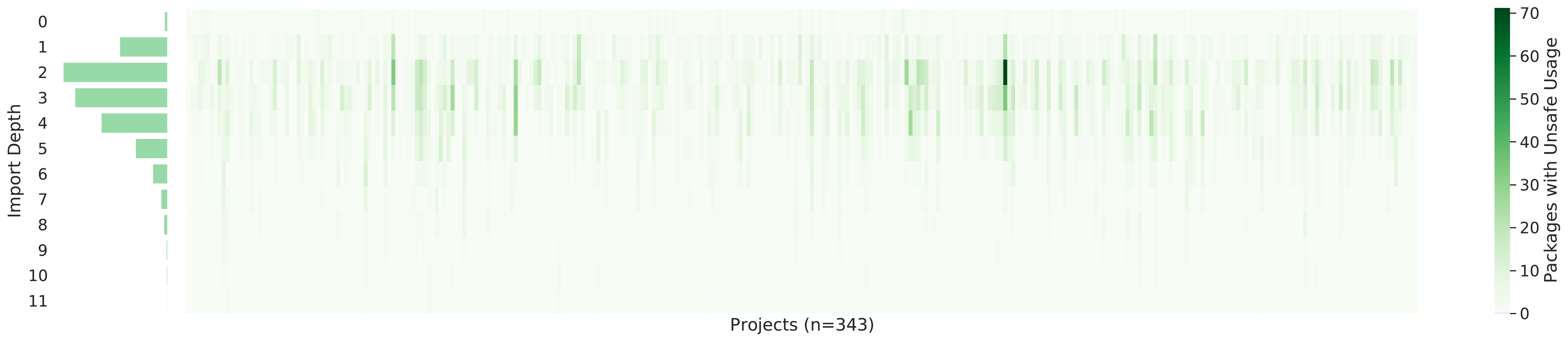}
    \caption{Import Depth of Unsafe Packages. Unsafe packages are around a depth of \averageUnsafeImportDepth{} (sd=\stdUnsafeImportDepth{})}
    \label{fig:unsafe-import-depth}
\end{figure*}

\subsection{Data Set}

As our research is focused on open-source projects, we crawled the \initalProjs{} most-starred Go projects available on GitHub. 
To further understand the influence of dependencies, we then selected the applications supporting \textit{go modules}.
With the introduction of Go \checkNum{1.13}, \textit{go modules}\footnote{\url{https://blog.golang.org/using-go-modules}} are the official way to include dependencies.
Unfortunately, \withoutModules{} of the projects did not yet support Go modules.
Thus, we excluded them from our set.
Furthermore, \notCompiled{} projects that did not compile were also removed.
As a result, we ended up with \projsAnalyzed{} top-rated Go projects. %
These have between 72,988 and 3,075 stars, with an average of 7,860. %

\subsection{Unsafe Usages in Projects and Dependencies}
\label{sec:eval:unsafewild}

We used the Go tool chain to identify the root module of each project. 
This is the module defined by the top-level \textit{go.mod} file in the project.
Then we enumerated the dependencies of the project, and built the dependency tree.
For each package, we used \toolUsage{} to generate CSV reports of the found \unsafe{} usages.
Through these analyses we answer the research questions of how many projects use \unsafe{} either in their own code or dependencies (\ref{rq:prevalApp}), and how deep in the dependency tree are the most \unsafe{} code usages (\ref{rq:depsDepth}). 
By selecting only results from the project root modules, we can easily find out how many applications contain a first-hand use of \unsafe{} code.
Our data shows that \checkNum{131} (\checkNum{38.19\%}) projects have at least one \unsafe{} usage within the project code itself.
By looking closer at the imported packages, we see that \checkNum{3,388} of \checkNum{62,025} (\checkNum{5.46\%}) transitively imported packages use \unsafe{}. 
There are \checkNum{312} (\checkNum{90.96\%}) projects that have at least one non-standard-library dependency with \unsafe{} usages somewhere in their dependency tree.
Since all projects include the Go runtime, which uses \unsafe{}, counting it as an \unsafe{} dependency would mean that \checkNum{100\%} of projects transitively include \unsafe{}.
We consider this to be less meaningful, as we assume the Go standard library is well audited and safer to use.

\begin{tcolorbox}[boxsep=1pt, enlarge top by=5pt, title=Answer to \ref{rq:prevalApp}]
About \checkNum{38\%} of projects directly contain \unsafe{} usages.
Furthermore, about \checkNum{91\%} of projects transitively import at least one dependency that contains \unsafe{}.
\end{tcolorbox}

Figure~\ref{fig:unsafe-import-depth} shows the number of packages with at least one \unsafe{} usage by their depth in the dependency tree for every project on its own as a heatmap, alongside the distribution for all projects combined as bars on the left side.
It is evident that most packages with \unsafe{} are imported early in the dependency tree with an average depth of \averageUnsafeImportDepth{}~and a standard deviation of \stdUnsafeImportDepth{}.
This number is very similar to the overall average depth of imported packages (\averageGeneralImportDepth{}). %
While the packages containing \unsafe{} can be manually found and evaluated, this process requires significant resources to handle the increasing number of packages introduced through each dependency. 
For developers only the first level of dependencies, the ones they added themselves, are really obvious.
On this level, \levelOneImportedUnsafePackagesCount{} out of \levelOneImportedPackagesCount{} imported packages (\levelOneImportedUnsafePackagesShare{}) contain \unsafe{}.

\begin{tcolorbox}[boxsep=1pt, enlarge top by=5pt, title=Answer to \ref{rq:depsDepth}]
Most imported packages containing \unsafe{} usages are found around a depth of \checkNum{3} in the dependency tree.
\end{tcolorbox}

\subsection{Types and Purpose of Unsafe in Practice}
\label{sec:eval:labeledData}

\begin{figure}[!t]
    \vspace{-12pt}
    \centering
    \includegraphics[width=0.43\textwidth]{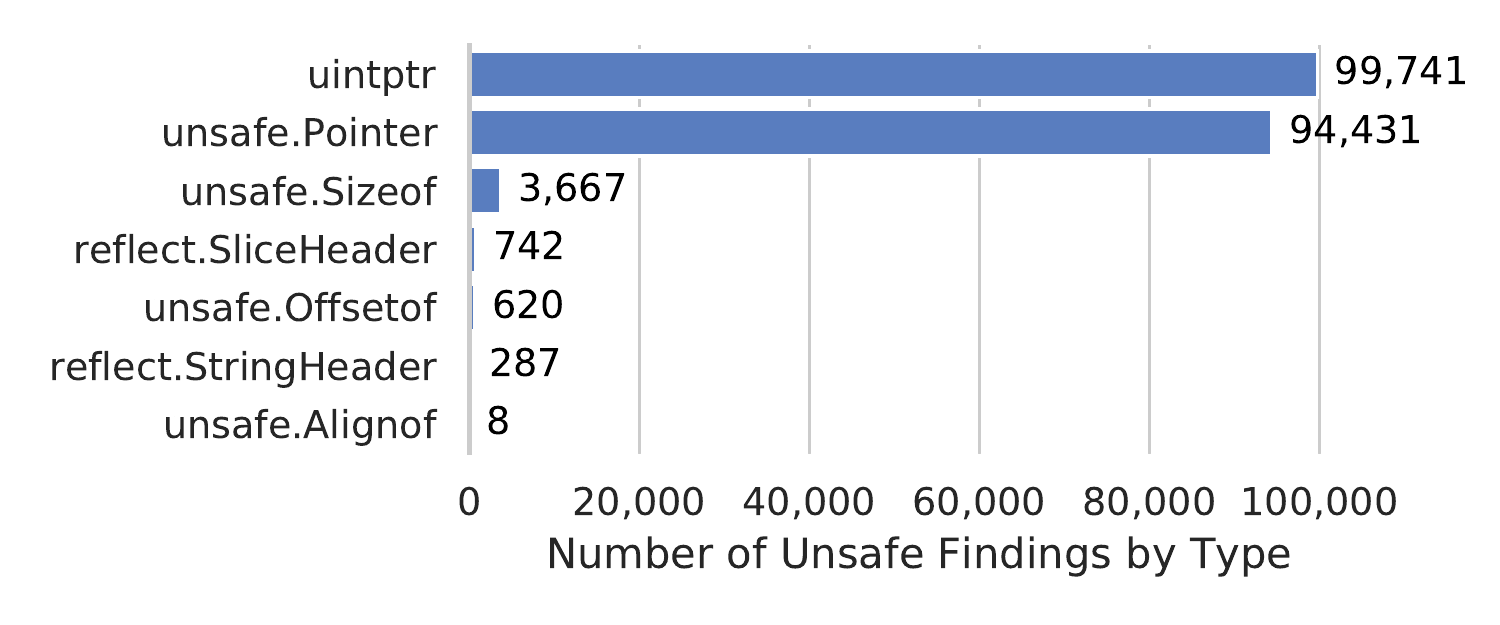}
    \caption{Distribution of different types of \unsafe{} tokens}
    \label{fig:unsafe-tokens-distribution}
\end{figure}

This section answers \ref{rq:distTypes} and \ref{rq:purpose}.
Figure~\ref{fig:unsafe-tokens-distribution} shows the distribution of the different \unsafe{} types in our data set.
Packages that are imported in different versions by the projects are counted once per version, as they might contain different \unsafe{} usages and coexist in the wild.
In our data set \textit{uintptr} and \textit{unsafe.Pointer} are used about equally often and are by far the most common with almost 100,000 findings. 
Next, \textit{unsafe.Sizeof} is still used a bit ($\sim 3,700$), while the other \unsafe{} types are rarely used~($< 1,000$).

\begin{tcolorbox}[boxsep=1pt, enlarge top by=5pt, title=Answer to \ref{rq:distTypes}]
In the wild, \textit{uintptr} and \textit{unsafe.Pointer} are orders of magnitude more common than other \unsafe{} usages.
\end{tcolorbox}

To learn about the purpose and context in which \unsafe{} is used, we needed to manually analyze code.
Thus, we selected the top \checkNum{10} projects (Table~\ref{tbl:dataset-projects}) with the most \unsafe{} usages in non-standard library packages.
From these projects and all their transitive dependencies, we randomly sampled \checkNum{400} code snippets that were found in the \textit{standard library (std)} and \checkNum{1,000} snippets from the remaining packages (\textit{app}).
We define standard library code as all packages that are part of the Go standard library or the \textit{golang.org/x/sys} module, as the \textit{syscall} standard library package is deprecated in favor of this module\footnote{\url{https://golang.org/pkg/syscall}}.
We split the snippets into two groups to analyze if there is a difference between the official standard library and non-standard library code regarding the usage of \unsafe{}.
Then, we identify class labels in two dimensions: what is being done, and for what purpose. 
Finally, we manually analyze all \checkNum{1,400} code snippets and label them accordingly.
The results of this process are shown in Table~\ref{tbl:dataset-classes}.

\begin{table}[!t]
    \vspace{1mm}

    \centering
    \caption{Projects selected for labeled data set}
    \label{tbl:dataset-projects}
    \begin{adjustbox}{max width=0.45\textwidth}
    \begin{tabular}{llrrl}
        {} & \textbf{Name} &  \textbf{Stars} &  \textbf{Forks} &    \textbf{Revision} \\ \hline
        \rowcolor{verylightgray}
        1  &         kubernetes/kubernetes &  66,512 &  23,806 &  \texttt{fb9e1946b0} \\
        2  &                 elastic/beats &   8,852 &   3,207 &  \texttt{df6f2169c5} \\
        \rowcolor{verylightgray}
        3  &             gorgonia/gorgonia &   3,373 &    301 &  \texttt{5fb5944d4a} \\
        4  &              weaveworks/scope &   4,354 &    554 &  \texttt{bf90d56f0c} \\
        \rowcolor{verylightgray}
        5  &  mattermost/mattermost-server &  18,277 &   4,157 &  \texttt{e83cc7357c} \\
        6  &               rancher/rancher &  14,344 &   1,758 &  \texttt{56a464049e} \\
        \rowcolor{verylightgray}
        7  &                 cilium/cilium &   5,501 &    626 &  \texttt{9b0ae85b5f} \\
        8  &                     rook/rook &   7,208 &   1,472 &  \texttt{ff90fa7098} \\
        \rowcolor{verylightgray}
        9  &             containers/libpod &   4,549 &    539 &  \texttt{e8818ced80} \\
        10 &                       xo/usql &   5,871 &    195 &  \texttt{bdff722f7b} \\ %
    \end{tabular}
    \end{adjustbox}
    \vspace{-10pt}
\end{table}
\begin{table*}[!t]
    \centering
    \caption[Labeled unsafe.Pointer usages in application code (non standard library) and standard library samples]%
    {Labeled unsafe.Pointer usages in application code (non standard library) and standard library samples \newline \footnotesize
        \underline{eff}: efficiency, \underline{ser}: (de)serialization, \underline{gen}: generics,
        \underline{no GC}: avoid garbage collection, \underline{atomic}: atomic operations,
        \underline{FFI}: foreign function interface, \underline{HE}: hide from escape analysis, \underline{layout}: memory layout control,
        \underline{types}: Go type system,
        \underline{reflect}: type reflection, \underline{unused}: declared but unused}
    \label{tbl:dataset-classes}
    \begin{adjustbox}{max width=\textwidth}
    
\begin{tabular}{r|cc|cc|cc|cc|cc|cc|cc|cc|cc|cc|cc|cc}
                    & \multicolumn{2}{c|}{\textbf{eff}} & \multicolumn{2}{c|}{\textbf{ser}} & \multicolumn{2}{c|}{\textbf{gen}} & \multicolumn{2}{c|}{\textbf{no GC}} & \multicolumn{2}{c|}{\textbf{atomic}} & \multicolumn{2}{c|}{\textbf{FFI}} & \multicolumn{2}{c|}{\textbf{HE}} & \multicolumn{2}{c|}{\textbf{layout}} & \multicolumn{2}{c|}{\textbf{types}} & \multicolumn{2}{c|}{\textbf{reflect}} & \multicolumn{2}{c|}{\textbf{unused}} & \multicolumn{2}{c}{\textbf{total}} \\ %
                    &  \textbf{app} &  \textbf{std} &  \textbf{app} &  \textbf{std} &  \textbf{app} &  \textbf{std} &   \textbf{app} &  \textbf{std} &    \textbf{app} &  \textbf{std} &  \textbf{app} &  \textbf{std} &  \textbf{app} &  \textbf{std} &    \textbf{app} &  \textbf{std} &   \textbf{app} &  \textbf{std} &     \textbf{app} &  \textbf{std} &    \textbf{app} &  \textbf{std} &   \textbf{app} &  \textbf{std} \\ \hline
                    
                    \textbf{cast} & 562 & 16 & 178 & 33 & 18 & & & & & & 24 & 6 && 2 & 3 & 13 & & 45 & 1 & & & & 786 & 115 \\ 
\rowcolor{verylightgray}
      \textbf{memory-access} &    2 &    1 &    9 &      &      &      &       &      &        &      &      &    1 &      &      &      4 &    6 &       &    4 &         &      &        &      &    15 &   12 \\
 \textbf{pointer-arithmetic} &    7 &    2 &    6 &    1 &      &      &       &      &        &    1 &      &    3 &    1 &    2 &      3 &    8 &       &    9 &         &      &        &      &    17 &   26 \\
\rowcolor{verylightgray}
         \textbf{definition} &    4 &    1 &   23 &      &    2 &      &       &      &        &      &    4 &    5 &      &      &        &    9 &       &    8 &       6 &    3 &        &      &    39 &   26 \\
           \textbf{delegate} &    4 &      &   64 &      &    2 &      &       &      &     11 &    5 &   29 &   45 &      &    4 &        &   14 &       &    6 &         &    1 &        &      &   110 &   75 \\
\rowcolor{verylightgray}
            \textbf{syscall} &      &      &      &      &      &      &    17 &  138 &        &      &      &      &      &      &        &      &       &      &         &      &        &      &    17 &  138 \\
             \textbf{unused} &      &      &      &      &      &      &       &      &        &      &      &      &      &      &        &      &       &      &         &      &     16 &    8 &    16 &    8 \\ \hline
                  \textbf{total} &  579 &   20 &  280 &   34 &   22 &    0 &    17 &  138 &     11 &    6 &   57 &   60 &    1 &    8 &     10 &   50 &     0 &   72 &       7 &    4 &     16 &    8 &  1000 &  400 \\
\end{tabular}

    \end{adjustbox}
        \vspace{-10pt}
\end{table*}

In the following, we outline the identified usage type classes describing what is being done in code.
The most prevalent are \textit{cast} operations from arbitrary types to other structs, basic Go types such as integers, slice/string headers, byte slices, or raw \textit{unsafe.Pointer} values. 
The \textit{memory-access} class is applied where \textit{unsafe.Pointer} values are dereferenced, used to manipulate corresponding memory or for comparison with another address.
\textit{Pointer-arithmetic} denotes usages of \unsafe{} to do some form of arithmetic manipulation of addresses, such as advancing an array.
\textit{Definition} groups usages where a field or method of type \textit{unsafe.Pointer} is declared for later usage.
\textit{Delegate} are instances where \unsafe{} is only needed in a function to pass it along to another function requiring a parameter of type \textit{unsafe.Pointer}. 
Thus, the need to use \unsafe{} is actually located elsewhere.
\textit{Syscall} are calls using the Go \textit{syscall} package or \textit{golang.org/x/sys} module.
As the name suggests, \textit{unused} is a class of occurrences that are not actually used in the analyzed code, e.g., dead code or unused parameters.

Our identified purpose classes, providing hints on why \unsafe{} is used, are described in the following.
\textit{Efficiency} includes cases where \unsafe{} is used only for the aim to improve time or space efficiency of the code.
The \textit{serialization} class contains (un)marshalling and (de)serialization operations such as in-place casts from complex types to bytes.
\textit{Generics} applies when \unsafe{} is used to build functionality that would otherwise be solved with generics if they were available in Go.
Samples in the \textit{avoid garbage collection} class are used to tell the Go compiler to not free a value while it is used, e.g., by a function written in assembly.
The \textit{atomic operations} class contains usages of the \textit{atomic} API which expects \unsafe{} for some functions.
The \textit{foreign function interface (FFI)} class contains interoperability with C code (CGo), and calling  functions that expect their parameters as \unsafe{} pointers.
\textit{Hide from escape analysis} includes the pattern described earlier (Listing~\ref{lst:unsafe-ex-escape-analysis}) to break the escape analysis chain.
The \textit{memory layout control} class contains code used for low-level memory management.
\textit{Types} snippets are used by the standard library to implement the Go type system.
\textit{Reflect} includes instances of type reflection and re-implementations of some types of the \textit{reflect} package, e.g., using \textit{unsafe.Pointer} instead of \textit{uintptr} for slice headers.
Again, \textit{unused} is a class of unused occurrences.

Using \unsafe{} for the sake of efficiency is the most prevalent motivation to use \unsafe{} in the wild covering \checkNum{58\%} in application code, whereas it is only used for this purpose in \checkNum{5\%} of the cases in std. 
From these, \checkNum{97\%} resp. \checkNum{80\%} are achieved by casting different types. 
The second biggest reason to use \unsafe{} in app is to perform some form of (de)serialization, accounting for \checkNum{28\%}.
For the standard library, the most relevant motivation is avoiding garbage collection with \checkNum{35\%}, whereas this is only used in \checkNum{2\%} of the usages in the app sample.
Furthermore, in std type \checkNum{18\%}, FFI \checkNum{15\%} and memory layout \checkNum{13\%} related \unsafe{} usages are rather common.
Both subsets share that hiding from escape analysis with \checkNum{0.1\%} (\textit{app}) and \checkNum{2\%} (\textit{std}) and using \unsafe{} for reflection with \checkNum{1\%} (both) are rare.
Implementation of generics functionality which is currently missing in Go is only done in few samples (\checkNum{2\%}), although some of the findings in the serialization class could alternatively be achieved with generics as well.

\begin{tcolorbox}[boxsep=1pt, enlarge top by=5pt, title=Answer to \ref{rq:purpose}]
\checkNum{More than half} of the \unsafe{} usages in projects and 3rd party libraries are to improve efficiency via \unsafe{} casts.
In the Go standard library, \checkNum{every third} use of \unsafe{} is to avoid garbage collection. 
\end{tcolorbox}

\subsection{Vulnerable Usages}
\label{sec:discussion}
Looking at the study results, we see that \unsafe{} is used consistently and wide-spread in the most popular open-source Go projects.
One might argue that the \new{usages}
found by \toolUsage{} are only minor annoyances, not severe or would require a manual case-to-case inspection. 
\new{Still, t}he exploitability of several of these %
\new{usages}
was discussed in \new{Section~\ref{sec:appr:vulnerabilites} with a reference to \checkNum{six} proof-of-concept exploits} that we developed\new{.}
\new{This c}learly \new{shows} that it is indeed possible to use the memory corruptions to one's advantage.
However, not all \unsafe{} usages contain a vulnerability. 
As already discussed, we implemented more specific checks for \new{two} %
patterns known to be problematic in \toolSA{} \new{(Section~\ref{sec:appr:toolSA})}.
\new{With it, three of the proof-of-concept exploits are mitigated, leaving the others which are much harder to detect statically for future work.}
The application of \toolSA{} to our data set revealed more than \numberBugsFixed{} \new{insecure usages of \unsafe{}}
in different projects.
Based on the results, we submitted so far \numberPRs{} pull requests to fix the\new{se usages.} 
By now, \numberPRsMerged{} have already been reviewed, acknowledged, and accepted by the corresponding project maintainers.
\section{Threats to Validity}
\label{sec:threatsToValidity}

Potential internal threats to validity for our study include bias towards bigger projects because those might be over-represented in the manually labeled data set. 
External threats include a bias towards more active projects with many developers because we selected a subset of the most-starred open-source projects on GitHub. 
Also we only considered projects that use the Go module system and about a third of the top 500 projects are not covered by the analysis yet.
Further, we could have missed projects from a special domain not having that many stars which might have other usage scenarios for \unsafe{} Go.
Nevertheless, one can argue that the biggest projects also have professional developers, higher standards and code gets more reviewed, thus, code quality should be higher.

\section{Related Work}
\label{sec:rw}

Previous research on Go mostly concentrated on issues related to its concurrency model including the channel implementation~\cite{tu2019,dilley2019,giunti2020,gabet2020,lange2017,bodden2016information}.
The work by Wang et al.~\cite{wang2020} suggests an improvement of the existing escape analysis in Go which we also discussed in our paper.

Moreover, the usage of \unsafe{} in other languages has already been studied to varying degrees. 
For Java, Mastrangelo et al.~\cite{mastrangelo2015} identified that 25\% of the analyzed artefacts depend on the Java \unsafe{} library.
The different JVM crash patterns caused by those usages are analyzed by Huang et al.~\cite{huang2019}.
Recently, two studies analyzed \unsafe{} usages in Rust projects and identified that \unsafe{} is widely used to improve performance or to reuse existing code~\cite{qin2020,evans2020}.
Furthermore, work was presented on how to ensure memory safety while using \unsafe{} in Rust~\cite{hussain2018Fidelius}.
Lehmann et al.~\cite{lehmann-everything-2020} studied to which extent \unsafe{} programs compiled to WebAssembly can lead to vulnerabilities within the virtual machine environment. %
For C/C++, non memory-safe languages, research exists on how to support at least partial memory safety~\cite{burow2018CUP, nagarkatte2009SoftBound} and work on identifying vulnerabilities by program analyses~\cite{song2019sok}.
A comprehensive study on memory-management-related vulnerabilities, like the ones we discussed earlier, and their mitigations is presented in earlier work~\cite{szekeres2013sok}.

Concerning project dependencies, it is difficult to count the dependencies that matter the most, e.g., by excluding test dependencies~\cite{pashchenko2018}.
A common problem is that dependencies are often updated slowly, keeping old bugs alive, although measures such as automated pull requests exist to mitigate this problem~\cite{derr2017keep, mirhosseini2017, lauinger2017}.

\section{Conclusion}
\label{sec:concl}

In this paper, we gave a systematic description of different dangerous programming patterns involving \unsafe{} and novel evidence on how to exploit these patterns. %
Furthermore, we presented two novel tools to help Go developers write safer code with respect to \unsafe{} Go and security analysts to evaluate \unsafe{} code.
First, \toolUsage{} identifies \unsafe{} usages not only within the main project package, but also in its transitive dependencies. 
Therefore it represents an effective tool to focus audit efforts on the code locations that are the most dangerous, raising awareness into how \unsafe{} is included into a project, and helps getting a general sense for the potential risks of deploying a specific project.
Second, \toolSA{} is a new static code analysis tool that helps developers identify dangerous code patterns that were previously uncaught with existing tools for linting.
Additionally, we conducted a study of \packagesAnalyzed{} packages from \projsAnalyzed{} top-starred open-source Go projects.
Here, we have shown that \unsafe{} is very common, especially when taking project dependencies into account. 
Finally, derived from this study, we presented a new data set of manually labeled code snippets, providing insight into how and for what purpose \unsafe{} is used by developers.
The reasons for introducing \unsafe{} operations are often tied to optimization, interoperability with external libraries or to circumvent language limitations.

In the future, supervised learning algorithms could use our labeled data set to train classifiers, which can then identify the purpose and domain of \unsafe{} usages by looking at new code. 
Furthermore, plugins for common IDEs that integrate our tools, \toolUsage{} and \toolSA{}, could be built to incorporate them into developers' workflow.%

\ifCLASSOPTIONcompsoc
\else
\fi

\section*{Acknowledgments}

This research work has been funded by the DFG as part of CRC 1119 CROSSING, by the German Federal Ministry of Education and Research and the Hessen State Ministry for Higher Education, Research and the Arts within their joint support of the National Research Center for Applied Cybersecurity ATHENE.

\bibliographystyle{IEEEtran}
\balance
\bibliography{IEEEabrv,references,references-withoutZotero}

\end{document}